\documentclass[12pt]{iopart}

\usepackage{iopams}
\usepackage{graphics,epsfig,subfigure}

\newcommand{\bin}[2]{C^{#2}_{#1}}
\newcommand{\nn}{\nonumber}
\newcommand{\bb}[0]{\begin{eqnarray}}
\newcommand{\ee}[0]{\end{eqnarray}}

\newcommand{\efig}[1]{Fig.~\ref{#1}}

\newcommand{\var}{\theta}

\newcommand{\ff}{\frac{1}{2}}

\renewcommand{\bin}[2]{C_{#1}^{#2}}

\newcommand{\dt}{\Delta t}

\begin{document}

\title{Dynamics of interval fragmentation and asymptotic distributions}

\author{Jean-Yves Fortin$^1$, Sophie Mantelli$^1$, and MooYoung Choi$^2$}

\address{$^1$CNRS, Institut Jean Lamour, D\'epartement de Physique de la
Mati\`ere et des Mat\'eriaux, UMR 7198, Vandoeuvre-les-Nancy, F-54506, France\\}
\address{$^2$Department of Physics and Center for Theoretical Physics,
Seoul National University, Seoul 151-747, Korea}
\ead{fortin@ijl.nancy-universite.fr,smantell@ijl.nancy-universite.fr,
mychoi@snu.ac.kr}
\date{\today}

\begin{abstract}
We study the general fragmentation process starting from one element of size
unity ($E=1$). At each elementary step, each existing element of size $E$ can be
fragmented into $k\,(\ge 2)$ elements with probability $p_k$.
From the continuous time evolution equation, the size distribution function $P(E;t)$
can be derived exactly in terms of the variable $z= -\log E$, with or
without a source term that produces with rate $r$ additional elements of unit size.
Different cases are probed, in particular when the probability of
breaking an element into $k$ elements follows a power law: $p_k\propto
k^{-1-\eta}$. The asymptotic behavior of $P(E;t)$ for small $E$ (or large $z$)
is determined according to the value of $\eta$. When $\eta>1$, the distribution
is asymptotically proportional to
$t^{1/4}\exp\left[\sqrt{-\alpha t\log E}\right][-\log E]^{-3/4}$ with $\alpha$ being a
positive constant, whereas for $\eta<1$
it is proportional to
$E^{\eta-1}t^{1/4}\exp\left[\sqrt{-\alpha t\log E}\right][-\log E]^{-3/4}$ with
additional time-dependent corrections that are evaluated accurately with the saddle-point
method.
\end{abstract}

%\keywords{}

\maketitle


\section{Introduction}

Numerous physical or social phenomena involve fragmentation processes, ranging from
fracture in geology, mass fragmentation of matter and stellar mass distributions
in astronomy \cite{brown83,grady88,brown89} to breakup of atomic nuclei, polymers
or colloidal matter~\cite{ziff86}, and replica symmetry breaking
of the Edwards-Anderson order parameter that determines more
accurately the ground state of spin glass systems in the framework of the
replica method~\cite{derrida87}. The fragmentation process depends greatly on
the distribution rate at which small elements are produced. Sequential models
\cite{brown89} for mass distribution of an aggregate use the production rate
starting from mass $m$ which is proportional to the fragmented mass $m'\,(<m)$
elevated to an adjustable negative power in the range $[-1,0]$. This will allow a
large number of small fragments to be produced, with a stationary mass
distribution belonging to the stretched exponential or Weibull class in the limit of small masses, on the assumption that there
exists a stationary self-consistent solution for the particle number
distribution. Such a skew distribution with a stretched exponent equal to 1/2 was also
found in a fragmentation process based on statistics~\cite{mott43,grady85}.
Weibull, log-normal, and other skew distributions emerge naturally from evolving systems~\cite{epl09,goh10} in
%other than fragmentation problems, such as
growth or multiplicative and
Yule~\cite{yule25} processes. This results from a master equation
describing the dynamics based on microscopic transition rates. Assuming a
power-law distribution for the size-dependent rate of fragmentation, with
exponent $\beta$, leads to scaling properties of the moments, which behave
non-uniformly as fragmentation is iterated. The exponents describing how moments
scale with time are dependent on implicit equations, and the dynamical
exponent of the size distribution function in the long-time limit is
given by $z_d=1/3$~\cite{ziff85} for binary fragmentation with no size
dependence of the fragmentation rate ($\beta =0$) and no fragments removed during the
process. More generally, we obtain $z_d= -[n(\beta-1)-1]^{-1}$ when $n$ fragments are
produced~\cite{hassan95}. % and $\beta$ non-zero.
Otherwise, when only $m \,(<n)$ fragments are kept in the process, the system possesses fractal properties with finite dimensions less than unity.
Other recursive and discrete fragmentation processes of intervals involve
binary fragmentation with probability $p$ and "freezing" of remaining
fragments (which do not break anymore) with probability
$1-p$~\cite{krapivsky00,dean02,krapivsky04}. This leads
to a stationary size-distribution which features a power-law solution
$E^{-2p}$ for size $E$. This still holds in higher dimensions. It also exhibits
critical behavior at the critical value $p_c=1/2$ above which the number of
fragments is infinite. % or finite otherwise.
The critical phenomena are analog to the Galton-Watson process for which branching is similar to fragmentation at different nodes~\cite{athreya}.

In this paper, we focus on the asymptotic size properties of
fragmented intervals and on the effects of
fragmentation rates. In particular, we are interested in
how these rates affect the long-time or small-size distributions in term
of dynamical processes in the continuous limit, which can be defined properly
from a discrete master equation. This is different from statistical or
stochastic fragmentation studied long ago~\cite{grady88,grady90} in massive
bodies made of a collection of small elements, where fragmentation is governed
by binomial statistics with a probability proportional to the fragmented mass,
and where fractures can occur at any point on the body independently of the
history of the previous fractures. Locations or distribution of these fractures
in this case satisfy a Poisson distribution and the cumulative distribution of
fragmented masses follows a simple exponential law.

Also, we will consider the presence of an external source, which
allows to study the typical time-dependent behavior of the distribution, by inserting at
regular time intervals an element of unit size ($E=1$) in the system.
The system, seen as a collection of different fragments, can therefore increase its total size, as its number of elements grows by successive breakings at a given rate. We propose
a standard description of the size distribution for the variable $z \equiv -\log E$
instead, which is suitable in this framework, and for specific fragmentation
rates depending on the size of the broken element. Unlike the general
fragmentation rate dependent only on the total size of
the element to be broken~\cite{ziff85}, we consider various possibilities of
breaking an element of size $E$ into $k\,(\ge 2)$ fragments, the $i$th of which has
the size $E_i$ satisfying $\sum_{i=1}^k E_i =E$.
%, with probability $p_k$.
Attention will be paid to the specific case that the rate of such fragmentation depends on $k$ and the corresponding probability $p_k$ satisfies a power-law distribution,
meaning that arbitrary small fragments can be produced with a controlled
parameter given by the exponent of the algebraic decay.

This paper is organized as follows:
In the second section, we first write the standard master equation for the size distribution $P(E;t)$ or $P(z;t)$ with $z\equiv -\log E$. %, with the continuous limit.
The third section presents a method of computing the moments, which is based on a generating function, with and without a source.
The fourth section is devoted to the saddle point analysis of general power-law quantities $p_k$, by means of the exact Fourier representation of $P(z;t)$. This
gives the small-size behavior ($E\ll 1$) of the distribution $P(E;t)$
with all corrective terms.
Finally, a summary is given.

\section{%Notations and
Evolution equation for binary fragmentation}

We consider an element of size unity ($E=1$) at time $t=0$. After one
iteration with time step $\dt$, this element is fragmented into two
pieces of arbitrary sizes with probability $p$ %($0\le p\le 1$),
or keeps its original size with probability $1-p$. We also consider the possibility of a
source which, after each iteration, produces an element of unit size ($E=1$)
with probability $p_s$. The distribution of elements of size $E$ at time $t$ is denoted by $P(E;t)$, with the initial %and normalized
condition $P(E;0)=\delta(E{-}1)$. The evolution of the size distribution $P(E;t)$ is
governed by the discrete evolution equation
\bb\fl
%\hspace{-1cm}
P(E; t{+}\dt)&=&(1-p)P(E;t)+2p\int_E^1\frac{P(E';t)}{E'}\,dE'
 - p P(E;t)+ p_s \delta(E{-}1),
 \label{discrete}
\ee
where the first term on the right-hand side describes the contribution from
the non-fragmentation process, the second term the contribution from
fragmentation of larger elements of size $E'\,(>E)$ that gives element of size $E$
(with the symmetry factor 2), the third term the (negative) contribution coming from the
fragmentation of element $E$ itself, and the last term corresponds to the source.
Introducing the fragmentation rate $\omega$ and the production rate $\omega_s$ in such a way that
$p =\omega \dt$ and $p_s =\omega_s \dt$ and taking the limit $\dt\rightarrow 0$, we
obtain \eref{discrete} in the continuum (dimensionless) form:
%We can take the continuum limit of the previous equation of motion, by
%considering the time limit $t:=2pT$ when $p\rightarrow 0$ and discrete time $T$
%large, the product being kept constant. In that case we obtain, after redefining
%the finite ratio $r\dt/(2p\dt)\rightarrow r$
%
\bb\label{eqmE}
\frac{\partial P(E;t)}{\partial t}
= -P(E;t)+\int_E^1\frac{P(E';t)}{E'}\,dE'+r\delta(E{-}1),
\ee
where time $t$ has been rescaled in units of $(2\omega)^{-1}$ and $r \equiv \omega_s/2\omega$ is the dimensionless production rate.

Integration of \eref{eqmE} over $E$ by parts leads in particular to $\frac{\partial}{\partial t}\int_0^1P(E;t)\,dE=r$,
which manifests that the net growth of the system comes from the production by the source term. It would be tempting to search for a power-law solution in the stationary regime,
%and $E\ne 1$,
i.e., $P(E;t)\propto E^{-\beta}$.
A short inspection shows that this kind of trial solution leads to the only possibility
$\beta=1$. Unfortunately, however, this would not provide a correct solution since
it is not normalizable due to the divergence near the origin.
It is clear in this case that the time $t$ plays an important role in the size
distribution function.

\subsection{Evaluation of the moments}

We now consider the variable $z=-\log E \,(\ge 0)$ instead of $E$,
so that the corresponding distribution is given by $P(z;t)=EP(E;t)$ or
$P(E;t)=\e^{z}\,P(z;t)$. In terms of the variable $z$, \eref{eqmE} reads
\bb
\label{eqmz}
\frac{\partial P(z;t)}{\partial t}
= -P(z;t)+\int_0^z dz'\,\e^{z'-z} P(z';t)+r\delta(z).
\ee
We then define the moments of $P(z;t)$ as
$M_n(t) \equiv \int_0^{\infty} dz\,z^n\,P(z;t)$ (up to a normalization factor), with the initial condition
$M_n(0)=\delta_{n,0}$ and in particular $M_0(t)=1+rt$.
The evolution equation for $M_n$ is given by
\bb\label{moments}
\frac{\partial M_n(t)}{\partial t}=-M_n(t) +\int_0^{\infty}\,dz\,z^n\e^{-z}
\int_0^z\e^{z'}P(z';t)\,dz'+r\delta_{n,0}.
\ee

The integral in \eref{moments} can be transformed, via integration by parts,
into $\int_0^{\infty} dz\,q_n(z) P(z;t)$, where we have set
$z^n\e^{-z} \equiv - d[\e^{-z}q_n(z)]/dz$ or $-q_n'(z)+q_n(z)=z^n$.
This equation bears simply the solution $q_n(z)=n!\sum_{k=0}^nz^k/k!$,
and accordingly, the evolution equation for the moments is given by
$\partial M_n/\partial t = n!\sum_{k=0}^{n-1}M_k/k!+r\delta_{n,0}$.
Here it is convenient to consider instead the quantities $R_n \equiv M_n/n!$,
which satisfy $\partial R_n/\partial t = \sum_{k=0}^{n-1}R_k+r\delta_{n,0}$
with the initial condition $R_n(0)=\delta_{n,0}$. In particular, we have
$R_0(t)=1+rt$, $R_1(t)=t+rt^2/2$, and $R_2(t)=t+(1+r)t^2/2+rt^3/6$.
Therefore the mean value of $z$ is given by $M_1=t$ and the variance by $\sigma^2=M_2-M_1^2=2t$.
To compute all other terms $R_n$, we introduce the generating function
$G(x;t) \equiv \sum_{n=0}^{\infty} x^n R_n$, which satisfies the differential equation
($x<1$)
\bb\label{Gdiff}
\frac{\partial G(x;t)}{\partial
t}=r+\frac{x}{1-x}G(x;t),
\ee
with the initial condition $G(x;0)=1$. It is then straightforward to obtain the solution
\bb\label{G}
G(x;t)=\exp\left(\frac{xt}{1-x}\right)-\frac{r(1-x)}{x}\left[1-\exp\left(\frac{xt}{1-x}\right)\right],
%G(x;t) = \left(1+\frac{r}{x}-r\right)\exp\left[\frac{xt}{1-x}\right] -\left(\frac{r}{x}-r\right),
\ee
from which all moments can be evaluated by successive differentiations.
This generating function is directly related to the characteristic function
of the distribution $P(z;t)$ as a function of time.
Namely, the time-dependent size distribution
is expressed in terms of the moments, through the Fourier transform:
\bb \label{PG}
P(z;t) = \int_{-\infty}^{\infty}\frac{d\lambda}{2\pi} \e^{i\lambda z}
           \sum_{n=0}^{\infty} \frac{(-i\lambda)^n}{n!} M_n
         = \int_{-\infty}^{\infty}\frac{d\lambda}{2\pi} \e^{i\lambda z} \,G(-i\lambda;t).
\ee

\subsection{No source: exact distribution}

Let us first consider the case of no source ($r=0$), where the generating function $G(x;t)$ in \eref{G} reduces to $G_0 (x;t) \equiv \exp[xt/(1-x)]$.
Equation (\ref{PG}), with $G(x;t)$ replaced by $G_0(x;t)$, leads to the integral representation
\bb\label{Przeroint}
P(z;t) = \e^{-t} \int_{-\infty}^{\infty}\frac{d\lambda}{2\pi}
\e^{i\lambda z} \exp\left[\frac{t}{1+i\lambda} \right].
\ee
Expanding the second exponential as a power series in $t$, we obtain
\bb
P(z;t)=  \e^{-t} \int_{-\infty}^{\infty}\frac{d\lambda}{2\pi}
\e^{i\lambda z} \sum_{n=0}^{\infty} \frac{(-it)^n}{n!}
\frac{1}{(\lambda-i)^n}.
\ee

Each term of the above series can be evaluated by means of the contour integration
on the complex $\lambda$-plane.
The term $n=0$ obviously gives the contribution $\e^{-t}\delta(z)$.
For all other terms ($n\ge 1$), carrying singularities at $\lambda=i$,
the residue theorem can be applied on the upper half plane Im\,$\lambda>0$,
to yield
\bb
P(z;t) = \e^{-t} \delta(z) + \frac{\e^{-t-z}}{z}\sum_{n=1}^{\infty}
\frac{(zt)^n}{n!(n-1)!}.
\ee
Identifying the series with the modified Bessel function of the first kind $I_1$:
$\sum_{n=1}^{\infty} x^n/[n!(n-1)!]=\sqrt{x}I_1(2\sqrt{x})$, we obtain
\bb\label{Przero}
P(z;t) = \e^{-t}\delta(z) + \e^{-t-z} \sqrt{\frac{t}{z}}\,I_1(2\sqrt{zt}).
\ee
Here normalization is directly satisfied by the fact that the Fourier
transform of $P(z;t)$, shown in \eref{Przeroint}, is unity when $\lambda=0$.
In \eref{Przero}, the first term proportional to the delta function corresponds to the remanent
portion of the initial element of size unity ($E=1$ or $z=0$), decreasing exponentially with time.
The second term describes the contributions of all fragmented intervals ($E<1$ or $z>0$)
after a finite time $t$, and becomes approximately equal to $t\e^{-t-z}$ when $t$ is small.

\subsection{Saddle point analysis}

It is interesting to compare the exact result in \eref{Przero} with the saddle
point of the argument function in \eref{Przeroint}, which is
\bb
\varphi=i\lambda z+\frac{t}{1+i\lambda},
\ee
in the regime of small intervals (or $z$ large).
It is convenient to let $\lambda \equiv iu$ with $u$ close to unity
on the upper-half complex plane. The saddle point value $u^*$ is determined
by the unique solution of $\varphi'(u^*)=0$ or $u^*=1-\sqrt{t/z}$.
This gives the main contribution of the argument $\varphi(u^*)=-z+2\sqrt{tz}$,
while the second derivative $\partial^2\varphi(u)/\partial u^2|_{u^*}=
2z^{3/2}/\sqrt{t}$ gives additional corrective terms. Performing a
Gaussian integration around the saddle point value, we obtain the asymptotic
distribution for large $z$:
\bb\label{PzeroAsympt}
P(z;t)&\approx& \frac{1}{2\sqrt{\pi}}
\frac{t^{1/4}}{z^{3/4}}\exp\left[-(\sqrt{t}-\sqrt{z})^2\right].
\ee
This result can also be obtained directly from \eref{Przero}, with the help of
the asymptotic form of the Bessel function: $I_1(x)\approx \e^{x}/\sqrt{2\pi x}$.
The distribution therefore decreases exponentially with $z$, in addition to
corrective terms in the argument of the form $\sqrt{z}$ and $\log z$.

%-------------------------------------------------------------------------------
 \begin{figure}
\centering
 \includegraphics[scale=0.5,angle=270,clip]{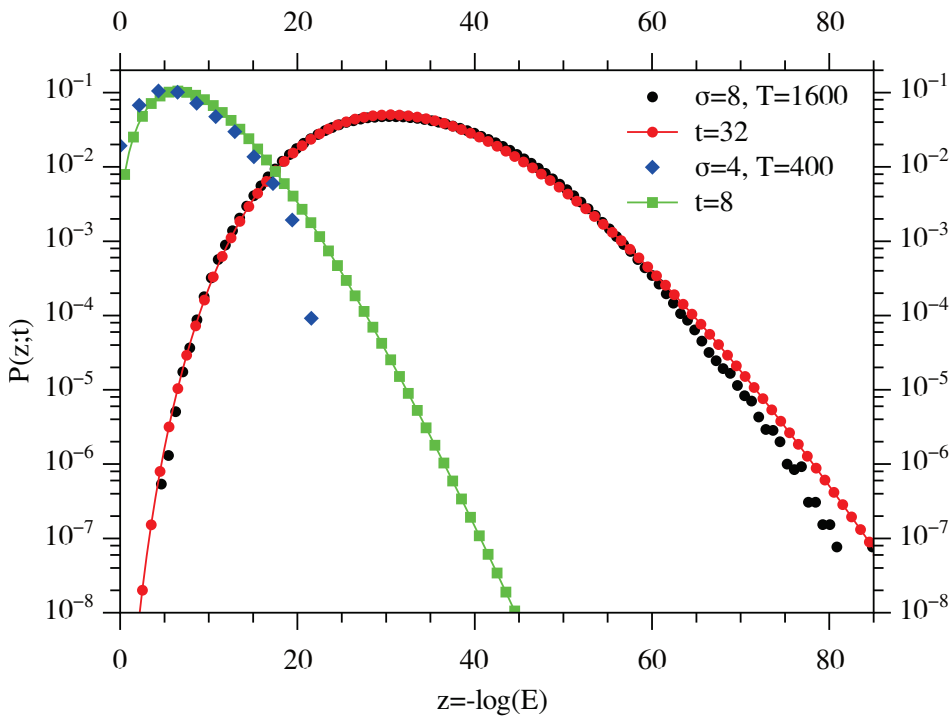}
 \caption{\label{figZero}Size distributions obtained numerically for $p=0.01$,
 after $T= 400$ iterations ($\sigma = 4$) (squares)
 and after $T=1600$ iterations ($\sigma=8$) (circles).
 Lines with symbols represent the corresponding analytical results in \eref{Przero}, at $t=8$ and $32$.}
 \end{figure}
%-------------------------------------------------------------------------------

Numerically, we performed simulations of the iterative process for give fragmentation probability $p$,
starting from a single element of unit size, until a sufficient number of fragments was produced in $T$ iterations,
%for given variance,
and computed their size distribution.
Since each iteration is performed in time interval $\dt$, the total elapsed time
is given by $T\dt$. Recalling that $t$ represents the dimensionless time,
measured in units of $(2\omega)^{-1}$, we thus have $T\dt = t/2\omega$ or
$t = 2pT$, relating $T$ and $t$.
Accordingly,
%the deviation $\sigma = \sqrt{2t}$ was evaluated and compared with the theoretical value $\sigma_T=\sqrt{4pT}$ after $T$ iterations (see text above).
the deviation after $T$ iterations takes the value $\sigma = \sqrt{4pT}$.
In \efig{figZero}, the distributions obtained numerically for two different values of
$T$ are compared with the analytical distribution in \eref{Przero} for the
corresponding time $t\,(=2pT)$
\footnote{One way in general of evaluating numerically the highly oscillating
integral in \eref{Przeroint} is to use a regularized
form for $z>0$, which avoids the oscillatory effects at large
$\lambda$ due to the Dirac peak located at
$z=0$: $P_{reg}(z;t)=P(z;t)-(2\pi)^{-1}\int_{-\infty}^{\infty}d\lambda
\e^{-t+i\lambda z}$}.
%
%Probability $p=0.01$ is taken small enough that continuous limit applies and curves are coincide
It is observed that the numerical data coincide with the analytical results
except for large values of $z$, where there arise deviations probably due to the limited number of
intervals in the region. Indeed such deviations tend to reduce as $T$ is increased.
Note that the mean number of elements produced in $T$ iterations is
given by $\langle n\rangle_T = (1+p)^T$ and the variance
$\langle n^2\rangle_T- \langle n\rangle_T^2 = (1-p)/(1+p)\langle n\rangle_T^2$ \cite{Jo11},
which indicates that for small $p$ the number of elements fluctuates strongly,
with the fluctuations of order of the number of elements produced. This can be explained with the
Galton-Watson theory~\cite{seneta69}, from the generating function
$g_T(z)=\sum_{n=1}^{2^T}z^nf_T(n)$, where $f_T(n)$ is the probability that at
time $T$ there are $n$ elements of various (indistinct) sizes. The generating
function satisfies the functional relation
$g_{T+1}(z)=(1-p)g_T(z)+pg^2_T(z)$, from
which we can deduce the moments corresponding to the number
of fragments produced randomly. For example, we have
$\langle n\rangle_T = \partial g_T(z)/\partial z|_{z=1}$.
Although the number of elements fluctuates strongly, the size distribution is convergent to the well-defined
distribution $P(z;t)$, even with only a small number of elements produced in a few time iterations, e.g.,
$\langle n\rangle_T \approx 53$ for $T=400$, as shown in \efig{figZero}.

\subsection{Presence of a source}

\begin{figure}
\centering
\includegraphics[scale=0.5,angle=270,clip]{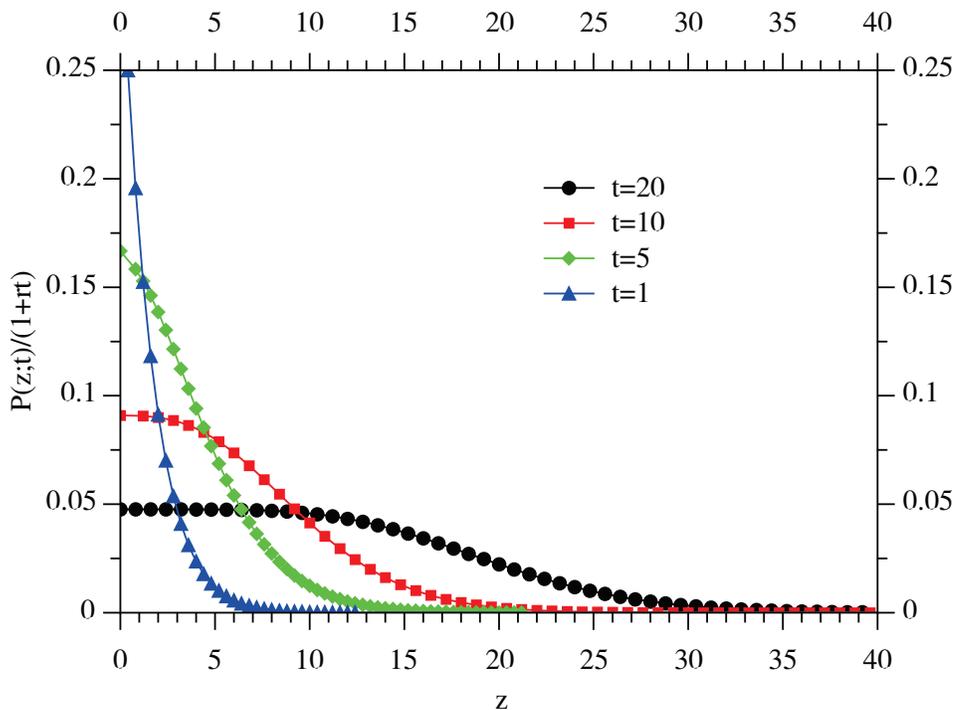}
\caption{\label{figPr} Size distribution normalized by the growth factor
$(1+rt)$, with the production rate $r=1$ at different times. The distribution becomes
flat for small sizes up to $z\approx t$, where the front is located.}
\end{figure}

When a source is present ($r\ne 0$), the size distribution can still be obtained from \eref{PG},
with the characteristic function $G(x;t)$ given by \eref{G}.
It is useful to notice that the time derivative of the term proportional to $r$, $G_1(x;t) \equiv G(x;t) - G_0(x;t)$,
is given by $\partial G_1(x;t)/\partial t = rG_0(x;t)$.
%\bb
% \frac{\partial}{\partial t} \left [-\frac{r(1-x)}{x}\left (1-\exp\left [\frac{xt}{1-x}\right ]\right )\right ]
%=r\exp\left [\frac{xt}{1-x}\right ]
%\ee
This reveals that the contribution of $G_1$ in \eref{PG} is equal to that of $G_0$ integrated over time $t$
and multiplied by $r$.
%then integrating over $x=-i\lambda$ gives the previous formula \eref{Przero} as
%function of the Bessel function $I_1$ before a second integration is performed
%over the time variable and initial condition $P(z;0)=\delta(z)$.
As a result, the size distribution is given by the sum of \eref{Przero} and its integral, and consists in
three terms:
\bb\label{Pr}\fl
P(z;t) &=& \e^{-t}\delta(z) + \e^{-t-z} \sqrt{\frac{t}{z}}\,I_1(2\sqrt{zt}) +
        + r\int_0^t d\tau\,\left[\e^{-t}\delta(z) + \e^{-z-\tau}
\sqrt{\frac{\tau}{z}}I_1(2\sqrt{z\tau})\right]
        \\ \nn\fl
 &=& \left[r+(1-r)\e^{-t}\right]\delta(z)+ \e^{-t-z}\sqrt{\frac{t}{z}}I_1(2\sqrt{zt})
  + \frac{r}{4z^2}\int_0^{2\sqrt{zt}}du\,u^2\,\exp\left(-z-\frac{u^2}{4z}\right) I_1(u),
\ee
where $\tau$ has been replaced by $u \equiv 2\sqrt{z\tau}$ in obtaining the last integral.
In the large time limit, the first term in \eref{Pr} coming from the source contributions reduces to $r\,\delta(z)$,
whereas the second term becomes negligible.
The resulting distributions for the production rate $r=1$ at time $t = 1, 5, 10,$ and $20$ are displayed in
\efig{figPr}. It is observed that a front wave develops with time, as the source introduces more unit size
elements in the system. This front is located approximately at $z \approx t$.

%-------------------------------------------------------------------------------
%-------------------------------------------------------------------------------

\section{General fragmentation rate}

We now consider the possibility that each interval element is allowed to be fragmented
into an arbitrary number of pieces, i.e., two (with probability $p_2$),
three (probability $p_3$), or $n$ (probability $p_n$) pieces,
with $\sum_{n=2}^{\infty} p_n=1$. % and $p_n$ arbitrary numbers.
In general, one can show that the generating function $G(x;t)$
satisfies a differential equation
$\partial G(x;t)/\partial t = Q(x)G(x;t)+r$,
with $Q(x)$ a rational function of $x$ having poles at positive integers
%poles greater or equal to one and located
on the real axis. The general evolution equation for $P(E;t)$ can be written
with the help of the probability $(n-1)(E'-E)^{n-2}/E'^{n-1}$ that an interval
of size $E'$ breaks into $n$ elements with one of given size $E\,(<E')$;
this may be confirmed by integrating successively over all possible sizes.
Note that this probability is uniform only for $n=2$.
Similarly to the case of binary fragmentation, the extended evolution equation
reads
\bb\label{eqmgenE}\fl
\frac{\partial
P(E;t)}{\partial t}
=-P(E;t)+\sum_{n=2}^{\infty}p_n(n-1)\int_E^1
\frac{(E'-E)^{n-2}}{E'^{n-1}}P(E';t)\,dE' + r\delta(E-1),
\ee
which, upon letting $E\equiv \e^{-z}$ and changing the variable from $E$ to $z$,
takes the form
\bb\fl\nn
\frac{\partial P(z;t)}{\partial t}&=&
- P(z;t)+ \sum_{m=2}^{\infty}p_m(m-1) \int_0^z dz'
\e^{-z+(m-1)z'}(\e^{-z'}-\e^{-z})^{m-2}P(z';t)
\\ \label{eqmgen}\fl
&+&r\delta(z).
\ee

Multiplying both sides by $z^n$ and integrating over $z$,
we obtain the evolution equations for the moments $M_n$ (or $R_n$).
Specifically, we expand the term $(\e^{-z'}-\e^{-z})^{m-2}$ by mean of the binomial
formula and perform the integration by parts,
% with the primitive integral
%$z^n\exp(-kz)\rightarrow -\sum_{l=0}^n\frac{n!}{l!k^{n-l+1}}z^l\exp(-kz)$.
to obtain
\bb \label{Rn}
\frac{\partial R_{n}(t)}{\partial t}=
-R_{n}(t)+\sum_{k=0}^{n} Q_{n-k} R_{k}(t)+r\delta_{n,0},
%\left[\sum_{m=2}^{\infty}(m-1)p_m
%\sum_{k=2}^{m-2}\frac{(-1)^k \bin{m-2}{k}}{(k+1)^{n-l+1}}\right].
\ee
where $Q_{n-k}\equiv \sum_{m=2}^{\infty}
(m-1)p_m\sum_{l=0}^{m-2} (-1)^l (l+1)^{-(n-k+1)} \bin{m-2}{l}$, and $Q_0=1$.
This differential equation can be solved with the help of the generating function
$G(x;t)$ defined previously.
Equation (\ref{Rn}) then leads to
\bb\label{Gn}
\frac{\partial G(x;t)}{\partial t}
=-G(x;t)+\sum_{n=0}^{\infty} x^n \sum_{k=0}^n Q_{n-k}R_k + r
= Q(x)G(x;t)+r,
\ee
where the double sum in $Q_{n-k}$ has been rearranged to give
\bb \label{Qdef}
Q(x) \equiv \sum_{n=0}^{\infty} x^n Q_n -1
= -1+\sum_{k=2}^{\infty}(k-1)p_k\sum_{l=0}^{k-2}\frac{(-1)^l\bin{k-2}{l}}{l+1-x}.
%\\
%&=&-1+\frac{p_2}{1-x}+2p_3\left [\frac{1}{1-x}-\frac{1}{2-x} \right ]
%+3p_4\left [\frac{1}{1-x}-\frac{2}{2-x}+\frac{1}{3-x}\right ]+\cdots
%=-1+\sum_{m=1}^{\infty}\frac{(-1)^{m-1}}{m-x}\sum_{k=m+1}^{\infty} (k-1)p_k \bin{k-2}{m-1}.
\ee

In the case of non-vanishing probabilities $p_k$ for $2 \leq k\leq n+1$,
$Q(x)$ is a rational function with simple poles located at $x=1,\cdots,n$.
Note that when only $p_2$ is non-zero, \eref{Gn}, together with \eref{Qdef},
reduces to \eref{Gdiff} for the binary fragmentation process.

As before, the size distribution function can be evaluated from \eref{PG}, where
the generating or characteristic function is given by the solution of \eref{Gn}:
\bb \label{Pgen}
G(x;t)= \e^{Q(x)t} - \frac{r}{Q(x)} \left[1-\e^{Q(x)t}\right].
\ee
In the limit $x\rightarrow 0$, we have $G(0;t)\rightarrow 1+rt$ and the system grows with
the scale factor $1{+}rt$, as expected.
Here $Q(x)$ can also be rewritten in terms of an integral by exponentiating $(l+1-x)^{-1}$
in \eref{Qdef} and performing the sum over $k$:
\bb
Q(x)=-1+\int_{0}^1d\,u\,u^{-x}\sum_{k= 2}^{\infty} (k-1)p_k (1-u)^{k-2}.
\ee
One can evaluate this integral precisely, performing successive integrations by parts,
particularly by differentiating $(1-u)^{k-2}$ and using the formula
$\int_0^1du\,(1-u)^ku^{-x}=k!\prod_{l=1}^{k+1}(l-x)^{-1}$.
We thus obtain
\bb \label{Qgen}
Q(x)
%&=&-1+\sum_{k\ge 2}p_k\frac{(k-1)!}{(1-x)(2-x)\cdots(k-1-x)} \\
=-1+\sum_{k=2}^{\infty} p_k\frac{\Gamma(1{-}x)\Gamma(k)}{\Gamma(k{-}x)}
=-1+\sum_{k=2}^{\infty}\frac{p_k}{\prod_{m=1}^{k-1}(1-x/m)},
\ee
which shows that %In the previous expression,
the poles of $Q(-i\lambda)$ are located on the imaginary axis, at values $\lambda=in$
with integer $n\ge 1$.

Deforming the integration path in \eref{PG} near the first pole $\lambda=i$, we obtain
an estimate for the saddle point value by writing $\lambda=iu$ with $u>0$ and
considering the series expansion of $Q$ near $u=1$:
$Q(u)\approx (1-u)^{-1}\sum_{k=2}^{\infty} (k-1)p_k$.
Then the argument function $\varphi(u)=-uz+tQ(u)$ (henceforth in the absence of a source) has a saddle point solution equal to $u^*=1-\sqrt{(\kappa-1) t/z}$,
where $\kappa=\sum_{k=2}^{\infty}kp_{k}$ is the mean number of fragments produced after breaking one interval element. When $\kappa$ is finite, we obtain the asymptotic solution of $P(z;t)$ for large $z$:
\bb \label{PgenAsympt}
P(z;t)&\approx&\frac{(\kappa{-}1)^{1/4}}{2\sqrt{\pi}}
\frac{t^{1/4}}{z^{3/4}}\exp\left[-z+2\sqrt{(\kappa{-}1)zt}-t \right].
\ee
Note that taking $p_2=1$ and $\kappa=2$, we recover \eref{PzeroAsympt} for binary fragmentation.
%
%In the case that $\kappa$ is infinite, on the other hand, the previous analysis is not valid anymore.

To be specific, we consider the case that the multiple fragmentation probabilities follow a power-law distribution:
$p_k = [\zeta(1{+}\eta)]^{-1}(k-1)^{-1-\eta}$
%corresponding to a multiple fragmentation process,
with $\eta$ positive and the zeta function $\zeta(1{+}\eta)$ giving the normalization factor. Then the mean number of fragments is given by $\kappa=1+\zeta(\eta)[\zeta(1{+}\eta)]^{-1}$, which is finite for $\eta>1$.
In this case, $Q(u)$ is approximated by
\bb
Q(u)\approx
-1+\frac{\zeta(\eta)}{\zeta(1{+}\eta)}\left[\gamma + \frac{1}{1-u}\right]
\ee
with the Euler constant $\gamma$, for which the saddle point value is given by
$u^*=1-\sqrt{\zeta(\eta)[\zeta(1{+}\eta)]^{-1}(t/z)}$.
This leads to the distribution in the form:
\bb
\label{PposAsympt}\fl
P(z;t)&\approx&
\frac{[\zeta(\eta)]^{1/4}}{2\sqrt{\pi}[\zeta(1{+}\eta)]^{1/4}}\frac{t^{1/4}}{z^{3/4}}\exp\left
[-z+2\sqrt{\frac{\zeta(\eta)}{\zeta(1{+}\eta)}zt}-t-\frac{\gamma\zeta(\eta)}{
\zeta(1{+}\eta)}t \right],
\ee
which is valid for $\eta >1$.
%This asymptotic formula is verified with numerical integration of \eref{Pgen}, displayed in \efig{figpkM} for $\eta=2$ and $t=4$.
The dominant contribution is given by the exponential decay $\e^{-z}$, similarly to the
previous binary fragmentation process.

When $0<\eta<1$, on the other hand, $\kappa$ becomes infinite, invalidating \eref{PposAsympt} based on \eref{PgenAsympt}. In this case it is convenient to rewrite \eref{Qgen} as
\bb
Q(u)=-1+\frac{1}{\zeta(1{+}\eta)}\sum_{k=1}^{\infty}\frac{\exp[\var_k(u)]}{k^{1+\eta-u}}
\ee
with the argument
\bb
\var_k(u)= -\sum_{m=1}^k\left [\log\Big (1-\frac{u}{m}\Big )+\frac{u}{m} \right
]+u\left [\sum_{m=1}^k\frac{1}{m}-\log k \right ].
\ee
In the limit of large $k$, function $\var_k(u)$ approaches rapidly the
well-defined finite limit: $\var(u)=\log\Gamma(1{-}u)$. However, when
$u$ approaches the value $\eta$, the sum of $k^{-1-\eta+u}$ diverges
before $u$ reaches the pole of the $\Gamma$ function, $u=1$, changing the position
of the saddle point value. It is then convenient to approximate $Q$ near
$u=\eta$ by
\bb\fl
Q(u)\approx
-1+\frac{\Gamma(1{-}u)}{\zeta(1{+}\eta)}\zeta(1{+}\eta{-}u) \approx
 -1+\frac{\Gamma(1{-}u)}{\zeta(1{+}\eta)}\left[
\frac{1}{\eta-u}+\gamma -\gamma_1(\eta{-}u)\right],
\ee
where $\gamma_1$ is the Stieltjes constant.
The saddle-point solution $u^*$, obtained by deriving the argument function
$\varphi(u)$ and solving $\partial \varphi(u)/\partial u=0$, leads
accurately to the saddle-point value satisfied by the quadratic equation
\bb
\frac{\Gamma(1{-}\eta)t}{\zeta(1{-}\eta)(\eta-u^*)^2}=z-C_{\eta}t,
\ee
where
\bb
C_{\eta}=\frac{\Gamma(1{-}\eta)}{\zeta(1{-}\eta)}\left[
\gamma_1-\Psi(1{-}\eta)\gamma-\ff\Psi'(1{-}\eta)-\ff\Psi(1{-}\eta)^2 \right]
\ee
with $\Psi(x) \equiv d\log\Gamma(x)/dx$. There are two solutions and the correct one
corresponds to $u^*<\eta$ or
\bb
u^* \approx \eta-\sqrt{\frac{\Gamma(1{-}\eta)t}{\zeta(1{+}\eta)(z-C_{\eta}t)}}
\ee
since deforming the path of integration starting from the real $\lambda$-axis to
$\lambda=iu$ with $u>0$ is possible without crossing the singularity $u=\eta$
only if $u<\eta$ is satisfied. Expanding $\varphi(u)$ up to the second order in $u$
around $u^*$ and integrating the local Gaussian leads to the asymptotic estimate
of $P(z;t)$ in the limit of large $z$, with $C_{\eta}t$ neglected:
\bb\fl\nn
P(z;t)&\approx&
\frac{[\Gamma(1{-}\eta)]^{1/4}}{2\sqrt{\pi}[\zeta(1{+}\eta)]^{
1/4}}\frac{t^{1/4}}{z^{3/4}}
\exp\left[
-\eta z+2\sqrt{\frac{\Gamma(1{-}\eta)}{\zeta(1{+}\eta)}zt}-t\right .
\\ \fl
&+&\left .\frac{\Gamma(1{-}\eta)}{\zeta(1{+}\eta)}\left[\gamma+\Psi(1{-}\eta)
\right]t \right],
\label{PasymptL}
\ee
which is valid for $\eta <1$.

\begin{figure}
\centering
\includegraphics[scale=0.5,angle=0,clip]{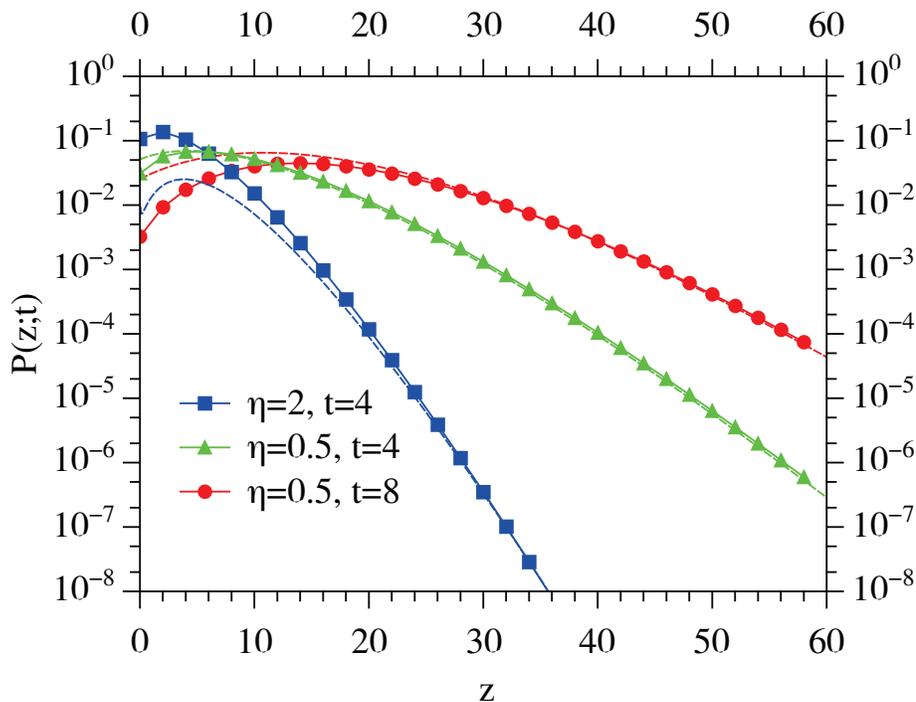}
\caption{\label{figpkM} Size distribution in the absence of a source with
the power-law distribution $p_k\propto k^{-1-\eta}$ of the number of elements
produced after one fragmentation. Dashed lines represent the asymptotic results based
on \eref{PasymptL} for $\eta<1$ and \eref{PposAsympt} for $\eta>1$,
respectively, by means of the saddle-point approximation and path deformation in the
integral of \eref{Pgen} near $\lambda=i$ for the first case and $\lambda=i\eta$ for
the second one.}
\end{figure}

Comparing \eref{PposAsympt} and \eref{PasymptL}, we notice that for $\eta< 1$ the
dominant exponential decay coefficient is governed by $\eta$ rather than unity in the case $\eta\ge 1$
for which the saddle-point solution is always given by $\lambda=i$ independently of $\eta$.
Corrective terms are proportional to $\sqrt{z}$ in the exponential argument,
with the difference that $\zeta(\eta)$ for $\eta>1$ is replaced by
$\Gamma(1{-}\eta)$ when $\eta<1$. Otherwise, the logarithmic correction has the
same factor $3/4$ in both cases.
The asymptotic formulae given by \eref{PposAsympt} and \eref{PasymptL} are verified with
numerical integration of \eref{Pgen}, as displayed in \efig{figpkM} for $\eta=2$ and $\eta=0.5$.

%-------------------------------------------------------------------------------
%-------------------------------------------------------------------------------

The case $\eta=1$ is special and needs a separate treatment. For this, we can express $Q(u)$ in terms of
the $\Gamma$ functions:
\bb
Q(u)=-1+\frac{6}{\pi^2}\left[\Psi'(-u)-\frac{1}{u^2} \right].
\ee
%where $\zeta(2)=\pi^2/6$. 
Near the singular value $u=1$, we can expand
the argument function $\varphi(u)$ and obtain the saddle-point solution:
$u^* \approx 1-[\pi^2z/(12t)-\zeta(3)]^{-1/3}$. It is clear that
$1-u^*$ scales as $z^{-1/3}$ instead of $z^{-1/2}$ and that
this solution gives a dominant term in $\e^{-z}$ and corrections which have
different behavior from the previous cases. Specifically, the size distribution takes the form
\bb\label{Pasympt1}
P(z;t) \approx
\frac{2^{1/6}}{3^{1/3}\pi^{5/6}}\frac{t^{1/6}}{z^{2/3}}
\exp\left[-z+\frac{3^{4/3}}{2^{1/3}\pi^{2/3}}z^{2/3}t^{1/3}\right]
\ee
for $\eta =1$.

\section{Discussion}

We have studied the general fragmentation process, in which each existing element of size $E$ can be
fragmented into $k$ elements with probability $p_k$.
The evolution equation for the size distribution function $P(E;t)$ has been built and solved
to yield $P(E;t)$ in the presence/absence of a source term producing elements of unit size.
Different cases have been probed, in particular when the probability of
breaking an element into $k$ elements follows a power law: $p_k\propto
k^{-1-\eta}$. The asymptotic behavior of $P(E;t)$ for small $E$ has been obtained according to the value of $\eta$.

In terms of the distribution $P(E;t)$ in the limit of small $E\,(\ll 1)$, the results are summaried as follows:
For $\eta>1$, the distribution is asymptotically given by
$P(E;t)\propto\exp[\sqrt{-\alpha t \log E}](-\log E)^{-3/4}$ with
$\alpha=4\zeta(\eta)/\zeta(1{+}\eta)$, whereas for $\eta<1$, we have
$P(E;t)\propto E^{\eta-1}\exp(\sqrt{-\alpha t\log E})(-\log E)^{-3/4}$ with
$\alpha=4\Gamma(1{-}\eta)/\zeta(1{+}\eta)$. For $\eta=1$, on the other hand,
we obtain $P(E;t)\propto\exp[(\alpha t\log^2 E)^{1/3}](-\log E)^{-2/3}$ with
$\alpha=27/[4\zeta(2)]$.
The asymptotic regime is thus dominated in general by whether $\eta$ is larger/smaller than unity or
whether the mean number $\kappa$ of fragments is finite. It also depends on
the location of the saddle point, relatively to the first pole of the
$\Gamma$ function, $\lambda=i$. In the special case $\eta=1$, we have obtained
the exact expression of the argument function $\varphi$ and treated accurately
the saddle-point value which differs from that in other cases. In view of classic
models of fragmentation, these results differ from
Moot-Linfoot~\cite{grady85,levy10} in the fact that variable $E$ in the
stretched exponential is replaced by $-\log E$ and that other corrections are present.
Generalization to more realistic cases is possible here, for example when
distributions $p_k$
depend statistically on the interval length $E$ and decrease as $E$ becomes smaller (i.e.
fragmentation becomes less effective when fragments are too small down to an intrinsic length of the system), and this can be studied using generating function \eref{Pgen} and \eref{Qgen}.

%\section*{acknowledgments}

\ack One of us (M.Y.C.) was supported by the National Research Foundation
through the BSR program (Grant Nos. 2009-0080791 and 2011-0012331).

\section*{References}
\bibliographystyle{iopart-num}
\bibliography{bib_frag}

\end{document}